\documentclass[footinbib,aps,pre,reprint,superscriptaddress,10pt]{revtex4-1}

\usepackage[osf,sc]{mathpazo}

\usepackage[scaled=0.90]{helvet}

\usepackage[scaled=0.85]{beramono}

\usepackage{graphicx}
\usepackage{graphics}
\usepackage{subfigure} 
\graphicspath{{./figures/}}

\usepackage{epsf,wrapfig,epsfig}
\usepackage[hidelinks]{hyperref}
\hypersetup{
	colorlinks=false,
	linkcolor=green,
	filecolor=cyan,      
	urlcolor=blue,
}

\usepackage{amsmath,amssymb}

\usepackage{color}

\makeatletter
\def\thefigure{\@arabic\c@figure}
\def\fps@figure{h, t}
\@addtoreset{equation}{section}

\makeatother

\begin{document}

\title{Experimental validation of phase space conduits of transition 
between 
potential wells}

\author{Shane D. Ross}
\author{Amir E. BozorgMagham} 
\author{Shibabrat Naik}
\email{Corresponding author: shiba@vt.edu}
\affiliation{Engineering Mechanics Program, Virginia Tech, Blacksburg, Virginia 
24061, USA}
\author{Lawrence N. Virgin}
\affiliation{Mechanical Engineering and Materials Science, Duke University, 
Durham, NC 27708, USA}

\begin{abstract} 
A phase space boundary between transition and non-transition trajectories, 
similar to those observed in Hamiltonian systems with rank one saddles, 
is verified experimentally in a macroscopic system. 
We present a validation of the  phase space flux 
across rank one saddles 
connecting adjacent potential wells
and confirm the underlying phase space conduits that mediate the transition. 
Experimental regions of transition are found to agree
with the theory to within 1\%, suggesting the robustness of phase space 
conduits of transition 
in a broad array of two or more degree of freedom 
experimental
systems, despite the presence of small dissipation.
\end{abstract}



\maketitle

\section{Introduction} 
Prediction of transition events and the determination of 
governing criteria has significance in many physical, chemical, and 
engineering systems where rank-1 saddles are present. To name but a few, 
ionization of a hydrogen atom under 
electromagnetic field in atomic physics~\cite{JaFaUz2000}, transport of 
defects in solid state and semiconductor physics~\cite{Eckhardt1995}, 
isomerization of clusters~\cite{Komatsuzaki2001}, reaction rates  in 
chemical physics~\cite{Komatsuzaki1999,WiWiJaUz2001}, buckling modes in 
structural mechanics~\cite{Collins2012,ZhViRo2018}, ship motion and 
capsize~\cite{Virgin1989,ThDe1996,NaRo2017}, escape and 
recapture of comets and asteroids in celestial 
mechanics~\cite{JaRoLoMaFaUz2002,DeJuLoMaPaPrRoTh2005,Ross2003}, and escape 
into inflation or re-collapse to singularity in 
cosmology~\cite{DeOliveira2002}.
The theoretical criteria of transition and its agreement with 
laboratory experiment have been shown for 1 degree of freedom (DOF) 
systems~\cite{Virgin1991, Gottwald1995, Novick2012}. 
Detailed experimental validation of the geometrical framework for 
predicting transition in higher dimensional phase 
space ($\geqslant 4$, that is for 2 or 
more DOF systems) is still lacking. 
The geometric framework of phase space conduits in such systems, termed tube 
dynamics\cite{KoLoMaRo2000,JaRoLoMaFaUz2002,DeJuLoMaPaPrRoTh2005,GaKoMaRoYa2006},
 has not before been demonstrated in 
a laboratory 
experiment. It is noted that  similar notions of transition were  
developed for idealized microscopic systems, particularly chemical 
reactions~\cite{Almeida1990,DeMeTo1991,JaFaUz2000,MaDe1989} under 
the names of transition state  and reactive island theory.
However, investigations of the 
 predicted phase space conduits of 
transition 
between wells in multi-well system have 
stayed within the confines of numerical simulations. 
In this paper, we present a direct experimental validation of the accuracy of 
the phase space conduits, as well as the transition 
fraction obtained as a function of energy, in a 4 dimensional phase space using 
a 
controlled laboratory experiment of a macroscopic system. 
\\
\indent
In~\cite{Baskan_2015,Baskan_2016,Figueroa_2017}, experimental validation of 
global
characteristics of 1 DOF Hamiltonian dynamics of scalar transport has been 
accomplished using direct measurement of the Poincar\'e stroboscopic sections 
using dye 
visualization of the 
fluid flow. In~\cite{Baskan_2015,Baskan_2016}, the experimental and 
computational 
results of chaotic mixing were compared by measuring the observed and simulated 
distribution of particles,  thus confirming the theory of chaotic transport 
in Hamiltonian systems for such systems. 
Our objective is to validate theoretical predictions of transition between 
potential wells in an exemplar experimental 2 DOF system, where qualitatively 
different global 
dynamics can occur.
Our  setup consists of a mass rolling on a multi-well surface that 
is representative of potential energy underlying systems that exhibit 
transition/escape behavior.
The archetypal potential energy surface chosen has implications in 
transition, escape, and recapture phenomena in many of the aforementioned 
physical systems. 
In some of these systems, transition in the conservative case has been 
understood in terms of trajectories of a given energy crossing a hypersurface 
or transition state (bounded by a normally hyperbolic invariant manifold of 
geometry $\mathbb{S}^{2N-3}$ in $N$ DOF). 
In this paper, for $N=2$, trajectories pass inside a tube-like separatrix, 
which has the advantage of accommodating the inclusion of 
non-conservative forces such as stochasticity and damping~\cite{NaRo2017,ZhViRo2018}. The semi-analytical geometry-based approach for identifying transition trajectories has also been considered for periodically forced 2 DOF systems in~\cite{gawlik_lagrangian_2009,onozaki_tube_2017}. Our analytical approach here focuses on identifying separatrices from the unforced dynamics, and generalizes to higher dimensional phase space \cite{WiWiJaUz2001,GaKoMaRo2005}.
Based on the illustrative nature of our 
laboratory experiment of a 2 DOF mechanical system, and the generality of the 
framework to higher degrees of freedom \cite{GaKoMaRoYa2006}, we envision the geometric 
approach demonstrated here can apply to experiments regarding transition across 
rank-1 saddles in 3 or more 
DOF systems in many physical contexts. 

\vspace{-3ex}
\section{Separatrices in N DOF}

To begin the mathematical description of the invariant manifolds that partition the 2N dimensional phase space, we perform a linear transformation of the underlying conservative Hamiltonian. This transformation involves a translation of the saddle equilibrium point to the origin and a linear change of coordinates that uses the eigenvectors of the linear system. The resulting Hamiltonian near the saddle has the quadratic (normal) form
\begin{equation}
\begin{aligned}
H_2(q_1, p_1, \ldots, q_N, p_N) = \lambda q_1 p_1 + \sum\limits_{k = 2}^N \frac{\omega_k}{2} \left( q_k^2 + p_k^2\right)
\end{aligned}
\end{equation}
where $N$ is the number of degrees of freedom, $\lambda$ is the real eigenvalue corresponding to the saddle coordinates (\emph{reactive coordinates} for chemical reactions) spanned by $(q_1, p_1)$ and $\omega_k$ are the frequencies associated with the center coordinates (\emph{bath coordinates} for chemical reactions) spanned by the pair $(q_k, p_k)$ for $k \in 2, \ldots, N$.

Next, by fixing the energy level to $h \in \mathbb{R}^+$ and $c \in \mathbb{R}^+$, we can define a co-dimension 1 region $\mathcal{R} \subset \mathbb{R}^{2N}$ in the full phase space by the conditions
\begin{equation}
H_2(q_1, p_2, \ldots, q_N, p_N) = h, \quad \text{and} \quad |p_1 - q_1| \leqslant c. 
\end{equation}
\begin{figure*}[!ht]
\centering
\includegraphics[width=0.95\textwidth]{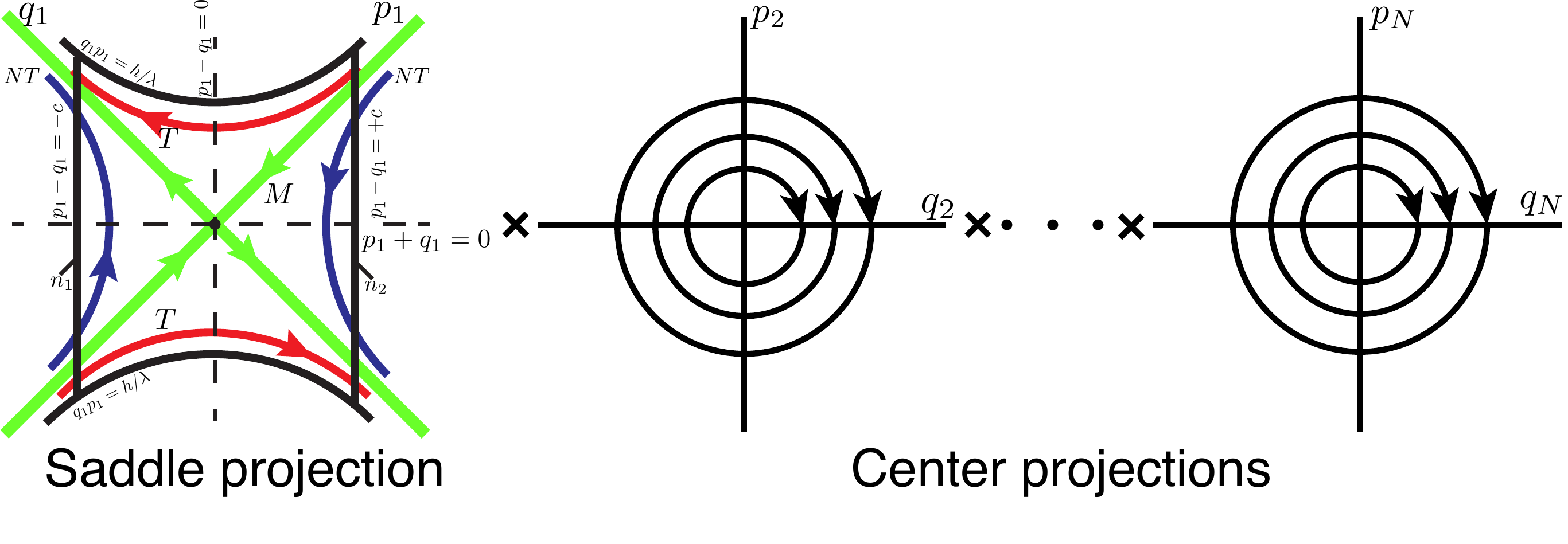}
\caption{The flow in the region $\mathcal{R}$ can be separated into saddle $\times$  center $\times \cdots \times$ center. On the left, the saddle projection is shown on the $(q_1,p_1)$-plane. The NHIM (black dot at the origin), the asymptotic orbits on the manifolds (M), two transition trajectories (T), and two non-transition trajectories (NT).
}
\label{fig:projection-saddle-center}
\end{figure*}
This implies that $\mathcal{R}$ is homeomorphic to the product of a $(2N - 2)$-sphere and an interval $I$, that is $\mathcal{R} \cong \mathcal{S}^{2N-2} \times I$ where the $\mathcal{S}^{2N-2}$ is given by
\begin{equation}
\frac{\lambda}{4} \left( q_1 + p_1 \right)^2 + \sum\limits_{k=2}^{N} \frac{\omega_k}{2}\left( q_k^2 + p_k^2 \right) = h + \frac{\lambda}{4}\left( p_1 - q_1 \right)^2.
\end{equation}
This bounding sphere of $\mathcal{R}$ at the middle of the equilibrium region where $p_1 - q_1 = 0$ is defined as follows
\begin{equation}
\mathcal{N}^{2N-2}_h = \left\{ (q,p) | \lambda p_1^2 + \sum\limits_{k=2}^N \frac{\omega_k}{2} (q_k^2 + p_k^2) = h \right\}, 
\end{equation}
corresponds to the transition state in chemical reactions (and other systems with similar Hamiltonian structure \cite{JaRoLoMaFaUz2002,NaRo2017,ZhViRo2018}). 

The following phase space structures and their geometry are relevant for understanding transition across the saddle:

a.~{\bf NHIM:} The point $q_1 = p_1 = 0$ corresponds to an invariant $(2N - 3)$-sphere, $\mathcal{M}^{2N-3}_h$, of periodic and quasi-periodic orbits in $\mathcal{R}$, and is given by 
\begin{equation}
\sum\limits_{k = 2}^N \frac{\omega_k}{2} \big( q_k^2 + p_k^2 \big) = h, \qquad q_1 = p_1 = 0.
\label{eqn:nhim_nDOF}
\end{equation}
This is known as the {\it normally hyperbolic invariant manifold} (NHIM) which has the property that the manifold has a ``saddle-like'' stability in directions transverse to the manifold and initial conditions on this surface evolve on it for $t \rightarrow \pm \infty$. The role of unstable periodic orbits the $4$ dimensional phase space (or more generally, the NHIM in the $2N$ dimensional phase space) in transition between potential wells is acting as anchor for constructing the separatrices of transit and non-transit trajectories. 

b.~{\bf Separatrix:} The four half open segments on the axes, $q_1 p_1 = 0$, correspond to four high-dimensional cylinders of orbits asymptotic to this invariant $\mathbb{S}^{2N - 3}$ either as time increases ($p_1 = 0$) or as time decreases ($q_1 = 0$). These are called {\it asymptotic} orbits and they form the stable and the unstable manifolds of $\mathbb{S}^{2N - 3}$. The stable manifolds, $\mathcal{W}_{\pm}^s(\mathbb{S}^{2N - 3})$, are given by
\begin{equation}
\sum\limits_{k = 2}^N \frac{\omega_k}{2} \big( q_k^2 + p_k^2 \big) = h, \qquad q_1 = 0.
\end{equation}
where $\pm$ denotes the left and right branches of the stable manifold attached to the NHIM. Similarly, unstable manifolds are constructed and are shown in the saddle space in Fig.~\ref{fig:projection-saddle-center} as four orbits labeled M. These form the ``spherical cylinders'' of orbits asymptotic to the invariant ($2N - 3$)-sphere. Topologically, both invariant manifolds have the structure of $(2N-2)$-dimensional ``tubes'' ($\mathbb{S}^{2N-3} \times \mathbb{R}$) inside the $(2N-1)$-dimensional energy surface. Thus, they separate two distinct types of motion: transit and non-transit trajectories. While a transition, passing from one region to another, trajectory  lies inside the ($2N - 2$)-dimensional manifold, the non-transition trajectories, bouncing back to their current region of motion, are those outside the manifold.

For a value of the energy just above that of the saddle, the nonlinear motion in the equilibrium region $\mathcal{R}$ is qualitatively the same as the linearized picture above~\cite{Moser1958,WiWiJaUz2001,Waalkens2010}. For example, the NHIM for the nonlinear system which corresponds to the $(2N - 3)$ sphere in~\eqref{eqn:nhim_nDOF} for the linearized system is given by 
\begin{widetext}
\begin{equation}
\begin{aligned}
\mathcal{M}_h^{2N - 3} = \Big\{ (q,p) \Big| \; \sum\limits_{k = 2}^N \frac{\omega_k}{2} \big( q_k^2 
+ p_k^2 \big) + f(q_2, p_2, \cdots, q_n, p_n) = h, \qquad q_1 = p_1 = 0. \Big\}
\end{aligned}
\label{eqn:NHIM_NF}
\end{equation}
\end{widetext}
where $f$ is at least of third order. Here, $(q_2, p_2, \cdots, q_N, p_N)$ are normal form coordinates and are related to the linearized coordinates via a near-identity transformation. In the neighborhood of the equilibrium point, since the higher order terms in $f$ are negligible compared to the second order terms, the $(2N - 3)$-sphere for the linear problem is a deformed sphere for the nonlinear problem. Moreover, since the NHIMs persist for higher energies, this deformed sphere $\mathcal{M}_h^{2N - 3}$ still has stable and unstable manifolds that are given by
\begin{widetext}
\begin{equation}
\begin{aligned}
\mathcal{W}_{\pm}^{S}(\mathcal{M}_h^{2N -3}) =& \Big\{ (q,p) \Big| \; \sum\limits_{k = 2}^N \frac{\omega_k}{2} \big( q_k^2 + p_k^2 \big) + f(q_2, p_2, \cdots, q_n, p_n) = h, \qquad q_1 = 0. \Big\} \\
\mathcal{W}_{\pm}^{u}(\mathcal{M}_h^{2N -3}) =& \Big\{ (q,p) \Big| \; \sum\limits_{k = 2}^N 
\frac{\omega_k}{2} \big( q_k^2 + p_k^2 \big) + f(q_2, p_2, \cdots, q_n, p_n) = h, \qquad p_1 = 0. 
\Big\}
\end{aligned}
\label{eqn:manifold_NF}
\end{equation}
\end{widetext}
This geometric insight is useful for developing numerical methods for {\it globalization} of the invariant manifolds using numerical continuation~\cite{Note1}. 

Now, we briefly describe the techniques that can be used to quantify and visualize the high 
dimensional invariant manifolds. For positive value of excess energy, one can use a normal form 
computation to obtain higher order terms of~\eqref{eqn:NHIM_NF} and~\eqref{eqn:manifold_NF}. A 
brief overview of this approach is given in~\cite{Wiggins2008, *Burbanks2008BackgroundAD} along 
with applications and results obtained using the computational tool for the Hamiltonian normal form. 
Another approach is to sample points on these manifolds since the geometry of the manifold is known 
near the equilibrium point. One would start by taking Poincar\'e sections and normal form theory 
that involves high-order expansions around a saddle $\times$ center $\cdots$ $\times$ center 
equilibrium. For example, in 3 DOF, the NHIM has topology $\mathbb{S}^3$ and thus a tube 
cross-section on a 4D Poincar\'e section will have topology $\mathbb{S}^3$ for which it is possible 
to obtain an inside and outside. If $x = {\rm constant}$ defines the Poincar\'e section, then one 
can project the $\mathbb{S}^3$ structure to two transverse planes, $(y,p_y)$ and $(z, p_z)$. On 
each plane, the projection appears as a disk, but because of the $\mathbb{S}^3$ topology, any point 
in the $(z, p_z)$ projection corresponds to a topological circle in the $(y,p_y)$ (and vice-versa) 
and from this, one can determine which initial conditions are inside, and thus transit 
trajectories, as has been performed previously \cite{GoKoLoMaMaRo2004,GaKoMaRo2005}. 


\section{Model of the 2 DOF experimental system}
\begin{figure}[!h]
	\vspace{-1ex}
	\centering
	\subfigure{\includegraphics[height=1.8in] 
		{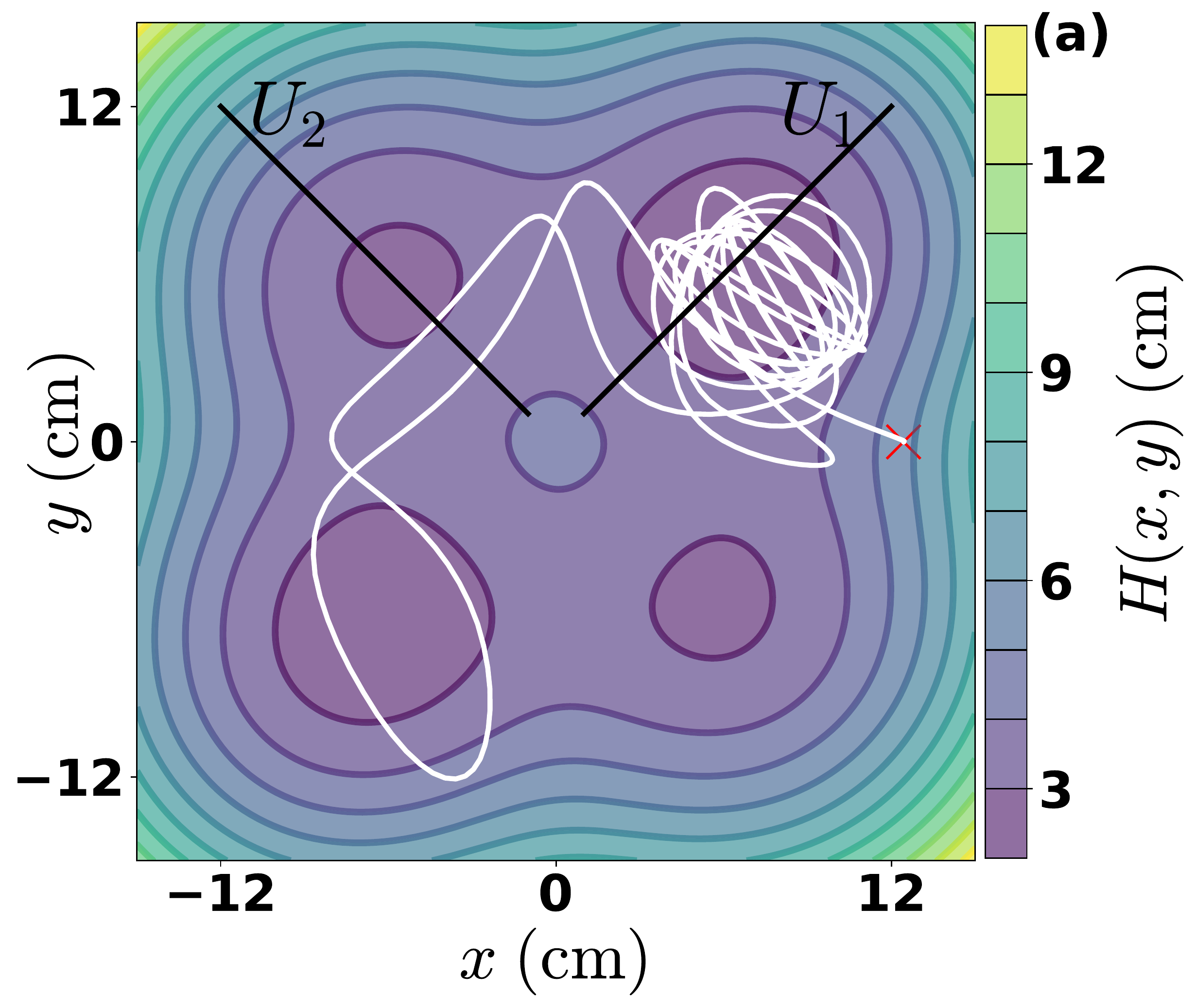}\label{fig:typ_traj}} %
	\subfigure{\includegraphics[height=1.8in]{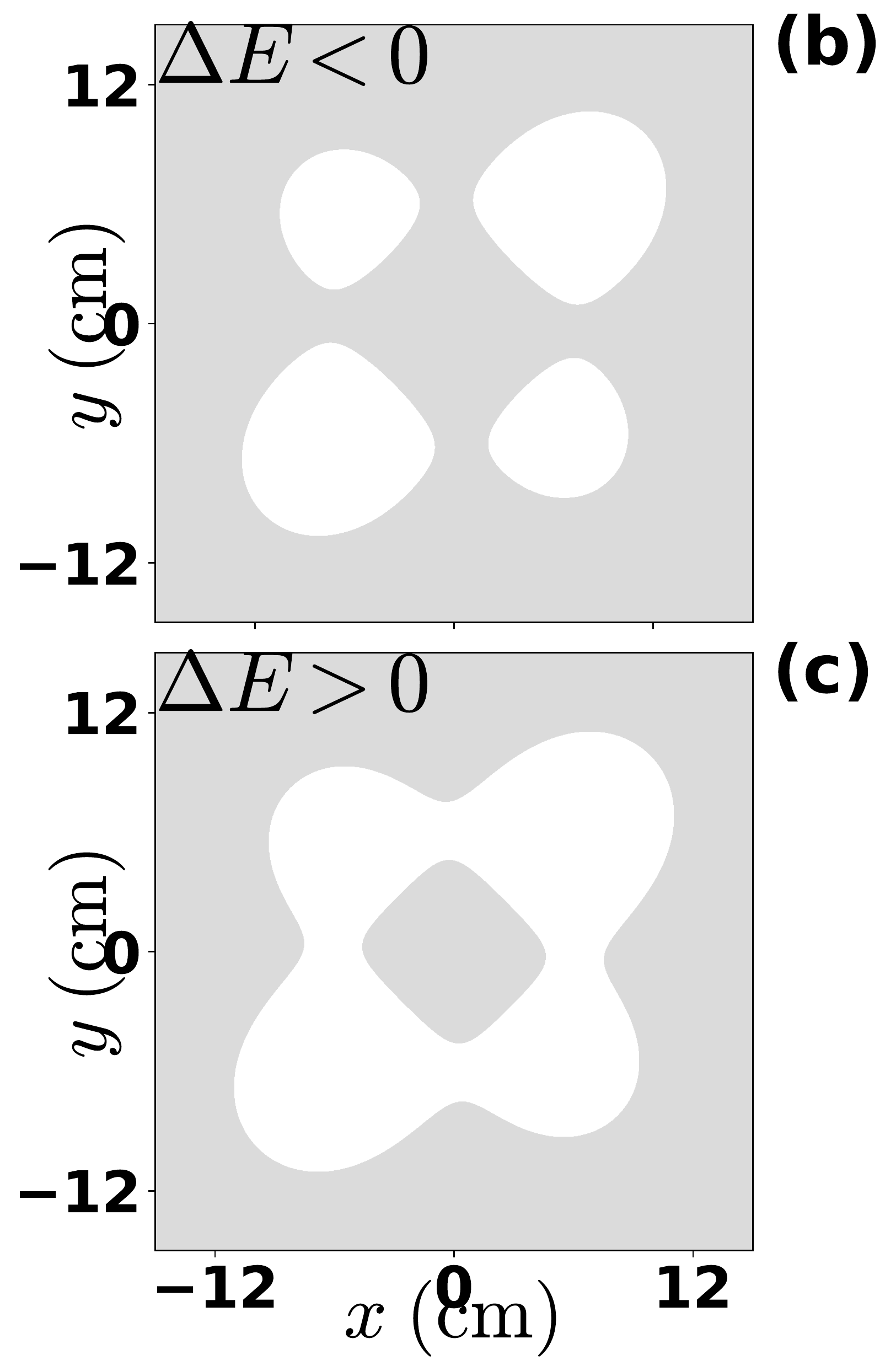}}
	\vspace{-2ex}
	\caption{\footnotesize{\textbf{(a)} A typical experimental 
	trajectory, shown in white, on the potential energy surface where 
	the contours denote isoheights of the surface. 
	This instance of the trajectory was traced by the ball released 
	from rest, marked by a red cross. \textbf{(b)} and 
	\textbf{(c)} 
	Show energetically accessible region projected on the configuration space 
	in white for $\Delta E < 0$: $\Delta E = -100 
	\;{\rm 	(cm/s)^2}$ and $\Delta 
	E > 0$: 
	$\Delta E = 100 \;{\rm (cm/s)^2}$, respectively.
	}}
	\label{fig:setup_surface_traj_sos}
	\vspace{-2ex}
\end{figure}

The initial mathematical model of the transition 
behavior of a rolling ball on the surface, $H(x,y)$, shown in 
Fig.~\ref{fig:setup_surface_traj_sos}, is described in~\cite{Virgin2010}.
The equations of motion are obtained from the Hamiltonian,
${\mathcal{H}}(x,y,p_x,p_y) = T(x,y,p_x,p_y) + V(x,y)$,  where mass factors out 
and
where the kinetic energy (translational and rotational for a ball 
rolling without slipping) is, 
{\small \vspace{-1ex}
\begin{equation}
\vspace{-1ex}
T = 
 \frac{5}{14} \frac{(1+H_y^2)p_x^2 + 
	(1+H_x^2)p_y^2 - 2 H_x H_y p_x p_y}{1 + H_x^2 + H_y^2}
\label{kin_rescale}
\vspace{-1ex}
\end{equation}}
\hspace{-1ex}where $H_{(\cdot)} = \frac{\partial H}{\partial 
(\cdot)}$. 
The potential energy is $V(x,y)=g H(x,y)$ where 
$g=981$ cm/s$^2$ is the gravitational
acceleration and the height function is
{\small
\vspace{-1ex}
\begin{align}
H = \alpha (x^2 + y^2) - \beta \left( \sqrt{x^2 + \gamma} + \sqrt{y^2 + 
	\gamma}  \right) -  \xi x y + H_0.
\label{pot_rescale}
\vspace{-3ex}
\end{align}}
%
\hspace{-1ex}This is the analytical function for the machined surface shown 
in~Fig.~\ref{fig:setup_surface_traj_sos}(b) and the isoheights shown 
in~Fig.~\ref{fig:setup_surface_traj_sos}(c).
We use parameter values $(\alpha,\beta,\gamma,\xi,H_0) = 
(0.07,1.017,15.103,0.00656,12.065)$ in the appropriate units~\footnote{See the 
Supplemental Material for derivation of equations 
of motion and the computational approach used to obtain the invariant 
manifolds.}. 

Let $\mathcal{M}(E)$ be the {\it energy manifold} in the 4D phase space given 
by setting 
the total energy 
  equal to a constant, $E$, i.e., $\mathcal{M}(E)=\{(x,y,p_x,p_y) 
\subset \mathbb{R}^4 \mid \mathcal{H}(x,y,p_x,p_y)=E\}$.
{
The projection of the energy manifold onto the $(x,y)$ configuration space 
 is the region of energetically possible motion for a mass with energy 
$E$,
and is given  by $M(E)=\{(x,y) \mid V(x,y)\leq E\}$. 
The boundary of $M(E)$ is  the zero velocity curve and is defined as the 
locus of points in the $(x,y)$ plane where the kinetic energy is zero. 
The mass is only able to move on the side of the curve where the kinetic energy 
is positive, shown as white regions in Fig.~\ref{fig:setup_surface_traj_sos}(d) 
and (e).
The critical energy for transition, $E_e$, is the energy of the 
rank-1 saddle  points in each bottleneck, which are all equal. This 
energy divides the global behavior of the mass into two cases, 
according to 
the sign of the excess energy above the saddle, $\Delta E = E-E_e$:

{\it Case 1:} $\Delta E < 0$ \textemdash the mass is safe against transition 
and remains inside the starting well since potential wells are not 
energetically connected (Fig.~\ref{fig:setup_surface_traj_sos}(d)). 

{\it Case 2:} $\Delta E > 0$ \textemdash the mass can transition by crossing 
the 
bottlenecks that open up around the saddle points, permitting transition 
between the potential wells (Fig.~\ref{fig:setup_surface_traj_sos}(e) and 
Fig.~\ref{fig:stable_tube_1-2_DelE100_trajs}(a) show this case). 

\begin{figure}[!t]
\vspace{-1ex}
\centering
\includegraphics[width=0.98\columnwidth]{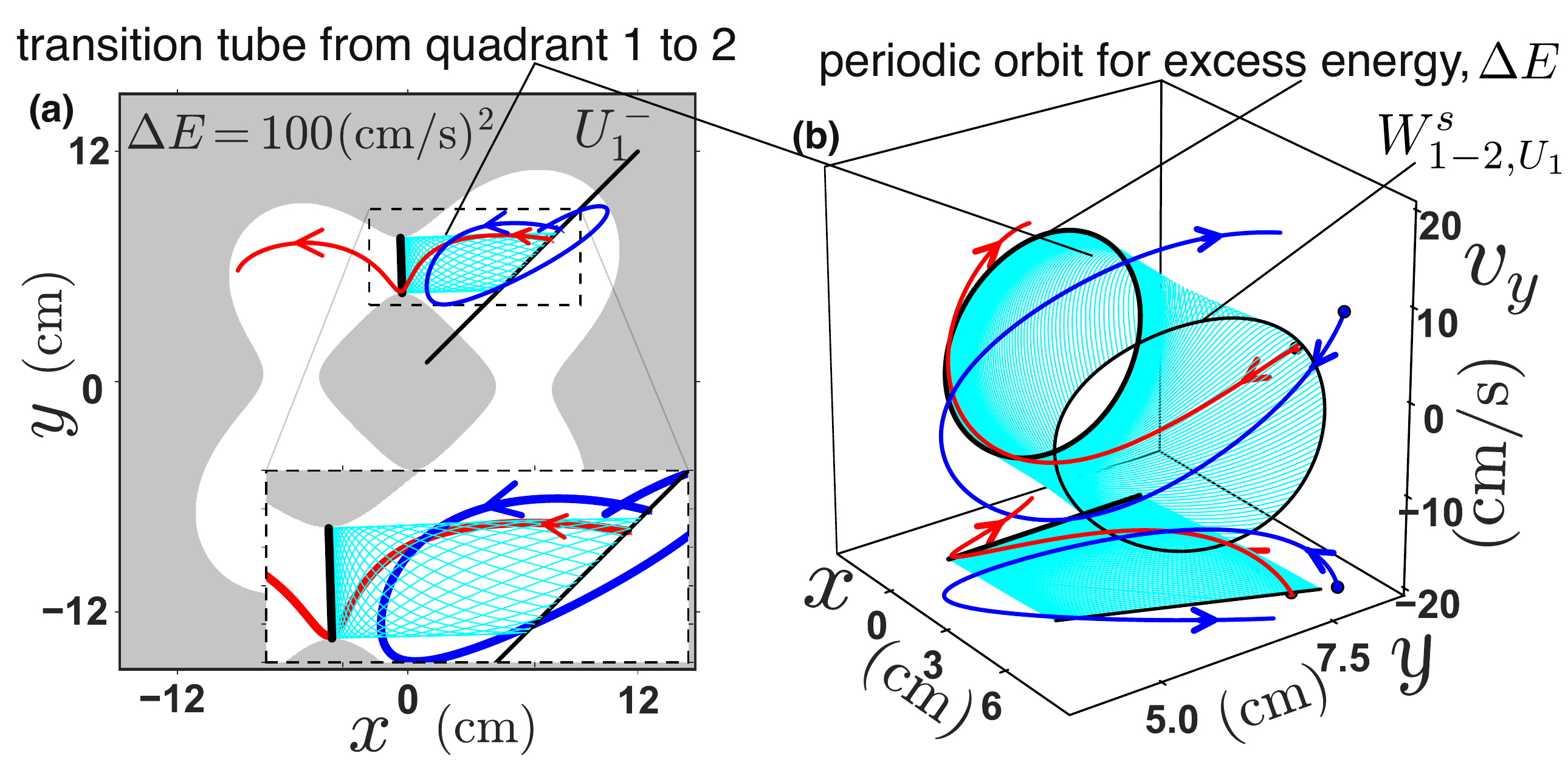}
\vspace{-2ex}
\caption{\footnotesize{(\textbf{a}) For a fixed excess energy, $\Delta E$, 
above the critical value $E_e$, the permissible regions (in white) are 
connected by a bottleneck around the saddle equilibria.  All motion from the 
well in quadrant 
1 to quadrant 2 must occur through the interior of a
stable manifold associated with an unstable periodic orbit in the 
bottleneck between the quadrants; seen as a 2D configuration space projection 
of the 3D energy manifold. We show the stable manifold (cyan) and the periodic 
orbit (black) for an excess energy of $\Delta E = 100 \;{\rm (cm/s)^2}$. A 
trajectory crossing the $U_1^-$ section inside the stable manifold 
will transition (red) into the quadrant 2 well, while one that is 
outside stays (blue) inside quadrant 1. The zoomed-in inset in the figure 
shows the structure of the manifold and how precisely the separatrix divides 
transition and non-transition trajectories.
\textbf{(b)} 
In the $(x,y,v_y)$ projection, the phase space 
conduit for imminent transition from quadrant 1 to 2 is 
the stable manifold (cyan) of geometry $\mathbb{R}^1 \times \mathbb{S}^1$ 
(i.e., a cylinder).
The same example
trajectories (red and blue) as in (a) that exhibit transition and 
non-transition behavior 
starting inside and outside the stable manifold, respectively, are shown 
in the 3D projection and projected on the ($x,y$) configuration space. 
A movie of a nested sequence of these manifolds can be found
\href{https://youtu.be/gMqrFX2JkLU}{here}.}}
\label{fig:stable_tube_1-2_DelE100_trajs}
\vspace{-3ex}
\end{figure}
Thus, transition between wells can occur when $\Delta E > 0$ and this 
constitutes a 
necessary condition. The sufficient condition for transition to occur is when a 
trajectory enters a codimension-1 invariant manifold associated with the 
unstable 
periodic orbit in the bottleneck as shown by non-transition and transition 
trajectories in 
Fig.\ref{fig:stable_tube_1-2_DelE100_trajs}(a)~\cite{KoLoMaRo2000}. In 2 
DOF systems, the periodic orbit residing in the bottleneck has an invariant 
manifold which is codimension-1 in the energy manifold and has topology
$\mathbb{R}^1 \times \mathbb{S}^1$, that is a cylinder or 
tube~\cite{Note1}. This implies that the transverse intersection of these 
manifolds with Poincar\'e surfaces-of-sections, $U_1$ and $U_2$, are 
topologically $\mathbb{S}^1$, a closed 
curve~\cite{KoLoMaRo2000,NaRo2017,ZhViRo2018}. 
All the trajectories transitioning to a different potential well (or having 
just transitioned into the well) are inside a tube manifold, for 
example as shown 
in Fig.~\ref{fig:stable_tube_1-2_DelE100_trajs}(b)~\cite{KoLoMaRo2000, 
GaKoMaRoYa2006}. 
For every $\Delta E > 0$, the tubes in phase space (or 
more precisely, within $\mathcal{M}(E)$) that lead to transition are the 
stable (and that lead to entry are the unstable) manifolds 
associated with the unstable periodic orbit of energy $E$. Thus, the mass's 
imminent 
transition 
between adjacent wells can be predicted by considering where it crosses 
$U_1$
as shown in Fig.~\ref{fig:sosU1pos}, relative to 
the intersection of the tube manifold. Furthermore, nested energy manifolds 
have corresponding nested 
stable and unstable manifolds that mediate transition.
To simplify analysis, we focus only on the 
transition of trajectories 
that intersect 
$U_1$ in the first quadrant. This surface-of-section is best described in polar 
coordinates $(r,\theta, 
p_r, p_{\theta}) $;
$U_1^{\pm } =  \{(r,p_r) ~|~ \theta = \frac{\pi}{4},~-{\rm sign}(p_{\theta}) 
=\pm 1\}$, 
where $+$ and $-$ denote motion to the right and left of the 
section, 
respectively
~\cite{Note1}. 
This Hamiltonian flow on $U_1^{\pm}$ defines a symplectic map with typical 
features such as  KAM tori and chaotic regions,  shown 
in Fig.~\ref{fig:sosU1pos} for two values of excess energy. 

Based on these phase space conduits that lead to transition, we would like to 
calculate what fraction of the energetically permissible 
trajectories will transition from/into a given well. 
This can be answered in part by calculating the transition rate of trajectories 
crossing 
the rank-1 saddle in the bottleneck connecting the wells. 
For computing this rate---surface integral of trajectories crossing a bounded 
surface per unit time---we  use the geometry of the tube manifold cross-section 
on the Poincar\'e section. For low excess energy, this computation is 
based on the theory of flux over a rank-1 saddle \cite{MacKay1990}, which 
corresponds to the action integral around the periodic orbit at energy $\Delta 
E$. By the 
Poincar\'e 
integral invariant~\cite{Meiss1992}, this 
action is preserved for symplectic maps, such as $P^\pm : U_1^\pm \rightarrow 
U_1^\pm$, and is equivalent to computing the area of the tube manifold's 
intersection with the surface-of-section. The transition fraction at each 
energy, $p_{\rm trans}(\Delta E)$, is calculated by the fraction of 
energetically permissible trajectories at a given excess energy, $\Delta E$, 
that will transition. This is given by the ratio of the cross-sections on $U_1$ 
of the tube to 
the energy surface. 
The transition area,
to leading order in $\Delta E$~\cite{MacKay1990}, is given by 
$A_{\rm trans} = T_{\rm po} \Delta E$, where $T_{\rm po} = 2 \pi / \omega$ is 
the 
period of the periodic orbits of small energy in the 
bottleneck, where $\omega$ is the imaginary part of the complex conjugate pair 
of eigenvalues 
resulting from the linearization about the saddle equilibrium 
point~\cite{MacKay1990}.
The area of the energy surface projection on $U_1$, to leading order in 
$\Delta E>0$, is $A_{\rm E} = A_{0} + \tau \Delta E $, where,
{\small 
\begin{align}
A_{0} =& 2 \int_{r_{\rm min}}^{r_{\rm max}} \sqrt{ 
\frac{14}{5}( E_e - gH(r))(1 + 4H_r^2(r)) } \;dr, \label{energy_area_U_2_crit} 
\\
\text{and} \; \tau =& \int_{r_{\rm min}}^{r_{\rm max}} \sqrt{ 
\frac{14}{5}\frac{ (1 + 4H_r^2(r))}{ (E_e - gH(r))}} \; dr. \; 
\label{energy_area_U_2}
\end{align}}
\hspace{-1ex}The transition fraction, under the well-mixed assumption mentioned 
earlier, is 
given in 2 DOF by 
{\small
\begin{equation}\label{transit_frac}
\begin{split}
p_{\rm trans} &= \frac{A_{\rm trans}}{A_{E}}  = \frac{T_{\rm 
po}}{A_{0}}  \Delta E \left( 1 - \frac{\tau}{A_{0}} \Delta E  + 
\mathcal{O}(\Delta E^2)\right).
\end{split}
\end{equation}}
\hspace{-1ex}For small positive excess energy, the predicted growth rate is 
$T_{\rm po}/A_0 \approx  0.87 \times 10^{-3} \; \rm (s/cm)^2$.
For larger values of $\Delta E$, the 
cross-sectional areas are computed 
numerically using Green's theorem, see 
Fig.~\ref{fig:energy_poincare_section}(b). 
\begin{figure}[!t]
	\vspace{-1ex}
	\centering
	\includegraphics[width=0.47\columnwidth]
		{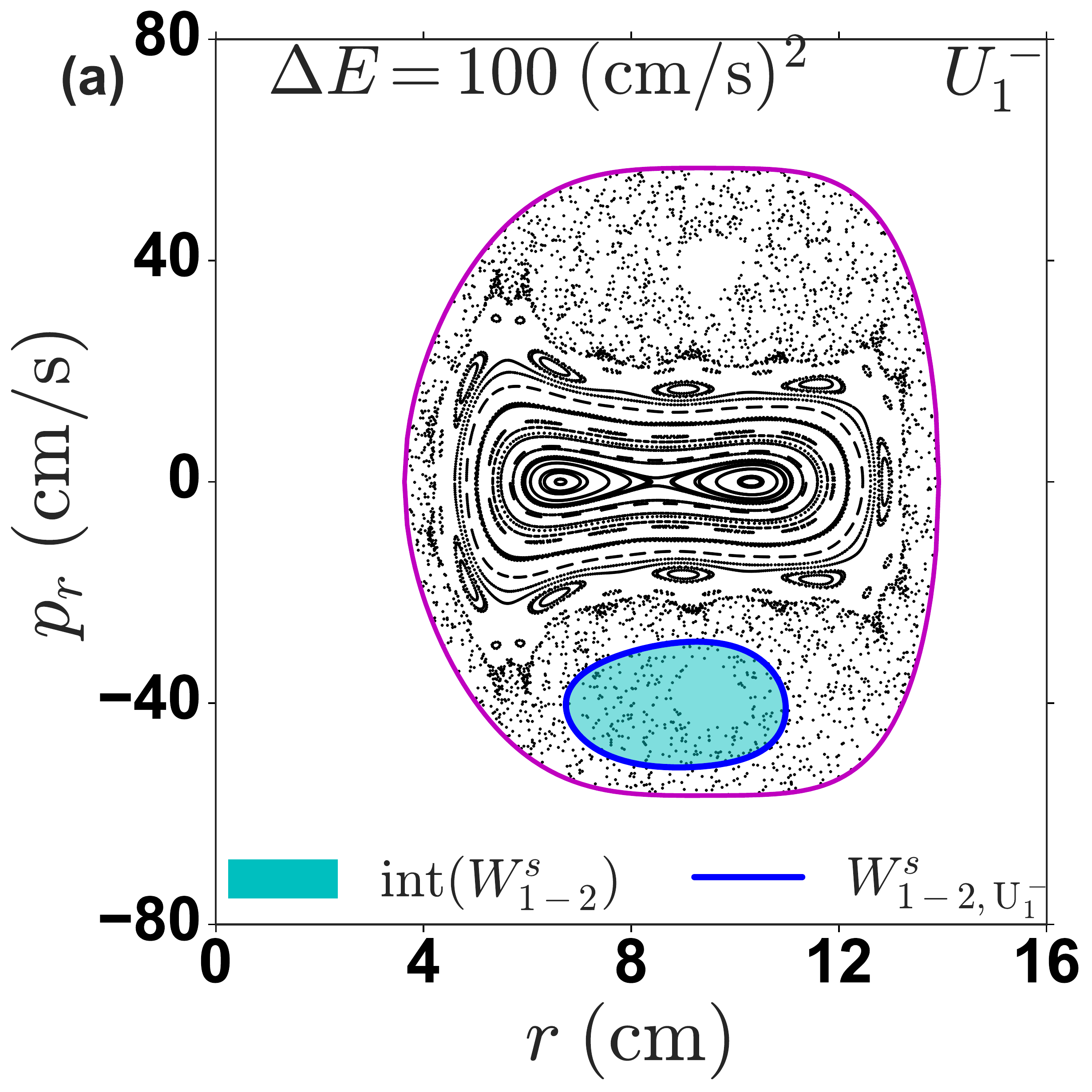}
		\label{sosU1m_DelE100_300pts_400finalT}
	\includegraphics[width=0.47\columnwidth]
	{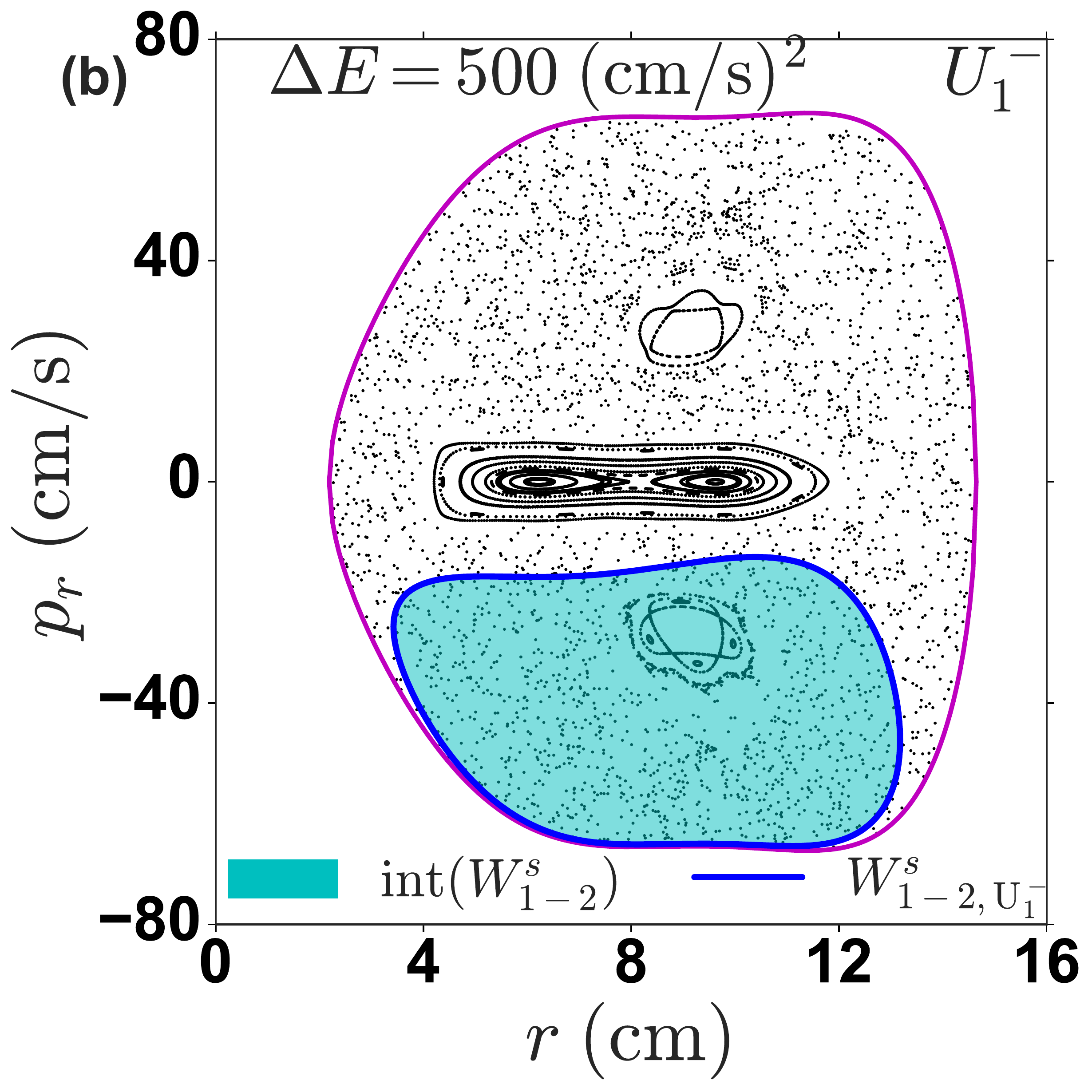}
	\label{sosU1m_DelE300_300pts_400finalT}
	\vspace{-3ex}
	\caption{\footnotesize{Poincar\'e section, $P^-: U_1^{-} 
	\rightarrow U_1^{-}$, of trajectories where $U_1^- := \{ (r, p_r) | \; 
	\theta = \pi/4, \; p_{\theta} > 0 \}$, at excess energy 
	\textbf{(a)} $\Delta E = 
	100 \;{\rm (cm/s)^2}$ and \textbf{(b)} $\Delta E = 
	500 \;{\rm (cm/s)^2}$.
	The blue curves 
	with cyan interior denote the intersection of the tube 
	manifold (stable) associated with the unstable periodic orbit with $U_1^-$. 
	It is to be noted that these manifolds act as a boundary between 
	transition and non-transition trajectories, and may include KAM 
	tori spanning more than one well. The interior of the manifolds, ${\rm 
	int(\cdot)}$, denote 
	the region of imminent transition to the quadrant 2 from quadrant 1. A 
	movie 
	showing the Poincar\'e section for a range of excess energy 
	can be found \href{https://youtu.be/sNvgXCrX6oo}{here}.}}
%
	\label{fig:sosU1pos}
	\vspace{-3ex}
\end{figure}

As with any physical experiment there is dissipation present, but over 
the time-scale of interest, the motion approximately conserves energy. 
We compare $\delta E$, the typical energy lost during a transition, with the 
typical excess energy, $\Delta E>0$, when transitions are possible. 
The  time-scale of interest, $t_{\rm trans}$, corresponds to the time between 
crossing $U_1$ and transitioning across the saddle into a neighboring well. The 
energy loss 
over  $t_{\rm trans}$  in terms of the measured damping ratio $\zeta \approx 
0.025$ is  $\delta E \approx \pi \zeta v^2(\Delta E)$ where the squared-velocity
$v^2(\Delta E)$ is approximated through the total energy. 
For our experimental trajectories, all starting at $\Delta E > 1000$ 
(cm/s)$^2$, we find $\delta E/\Delta E \ll 1$,
suggesting the appropriateness of the assumption of short-time conservative 
dynamics to study 
transition between wells~\cite{NaRo2017,ZhViRo2018}.
\vspace{-3ex}
\section{Experimental setup}
We designed a surface (shown in Fig.~\ref{fig:setup_surface_traj_sos}(b)) that 
has 4 wells, one in each quadrant, with saddles connecting the neighboring 
quadrants. 
The surface has 4 stable and 5 saddle (4 rank-1 and 1 rank-2) equilibrium 
points. Inter-well first order transitions are defined as crossing the rank-1 
saddles between the wells. 
On this high-precision machined surface, accurate to within $0.003~{\rm mm}$ 
and made using stock polycarbonate, a 
small rubber-coated spherical steel 
mass released from 
rest can 
roll without slipping under the influence of gravity. 
The mass is released from different locations on the 
machined 
surface to generate experimental trajectories. The mass is tracked using a 
Prosilica 
GC640 digital camera mounted on a rigid frame attached to the surface as shown 
in Fig.~\ref{fig:setup_surface_traj_sos}(a), with a pixel resolution of about 0.16 
cm. 
\begin{figure}[!h]
\vspace{-1ex}
\includegraphics[width=0.95\columnwidth]{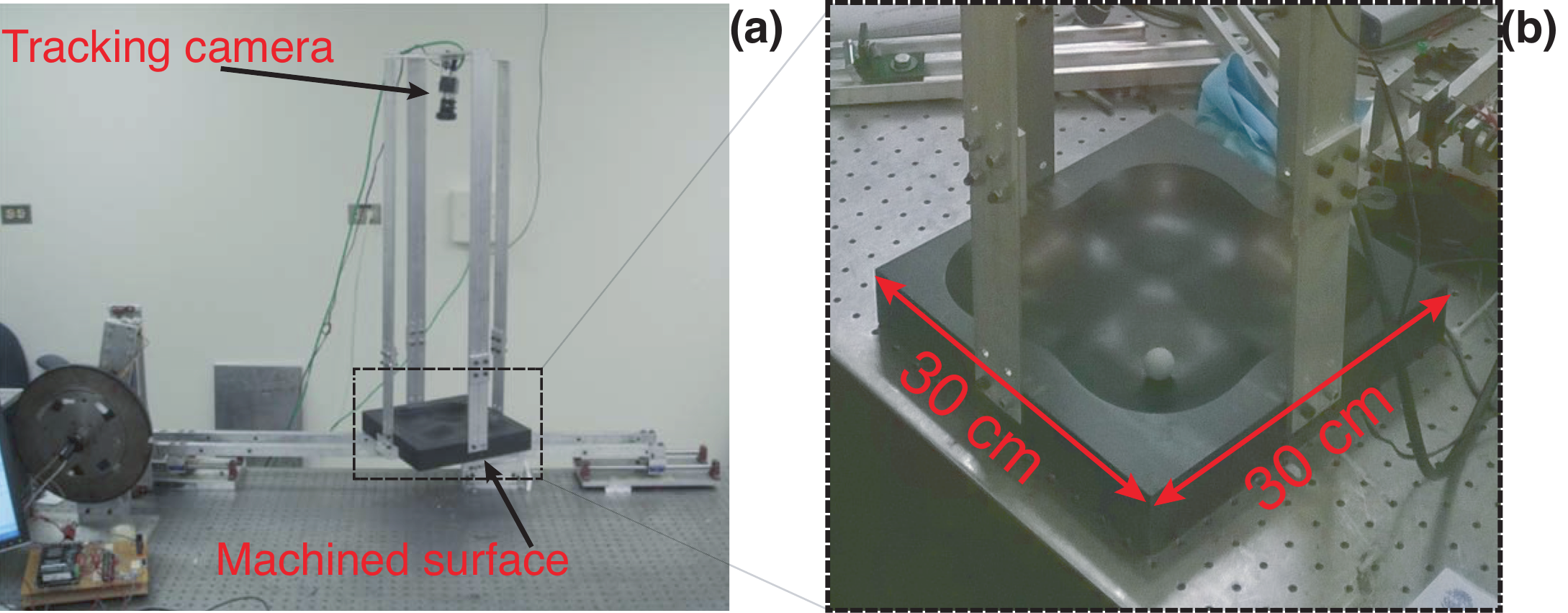}
\caption{\footnotesize{\textbf{(a)},~\textbf{(b)} Experimental apparatus showing the machined surface, tracking camera, and the rubber coated steel ball.}}
\label{fig:rolling_ball_expt}
\vspace{-2ex}
\end{figure}
The tracking is done by 
capturing black and white images at 50 
Hz, and calculating the coordinates of the mass's geometrical center. We 
recorded 120 experimental trajectories of about 10 seconds long, only using 
data 
after waiting at least the Lyapunov time of $\approx0.4$ 
seconds~\cite{Virgin2010} ensuring that the trajectories were 
well-mixed in the phase space. 
To analyze the fraction of trajectories that leave/enter a well, we obtain 
approximately
4000 intersections with a Poincar\'e surface-of-section, $U_1$, shown as a 
black 
line, 
for the analyzed range of energy. One such trajectory 
is shown in white in Fig.~\ref{fig:setup_surface_traj_sos}(c). 
These intersections are then sorted 
according to energy.  The intersection 
points on $U_1$ are classified as a transition from quadrant 1 to 2 if the 
trajectory, followed forward in time, leaves 
quadrant 1. Four hundred  transition events were recorded. 

\vspace{-3ex}
\section{Results}
For each of the recorded trajectories, we detect 
intersections with $U_1$ and 
determine the instantaneous $\Delta E$. 
\begin{figure}[!t]
\vspace{-1ex}
\begin{center}
\begin{tabular}{c}
\includegraphics[height=1.6in]{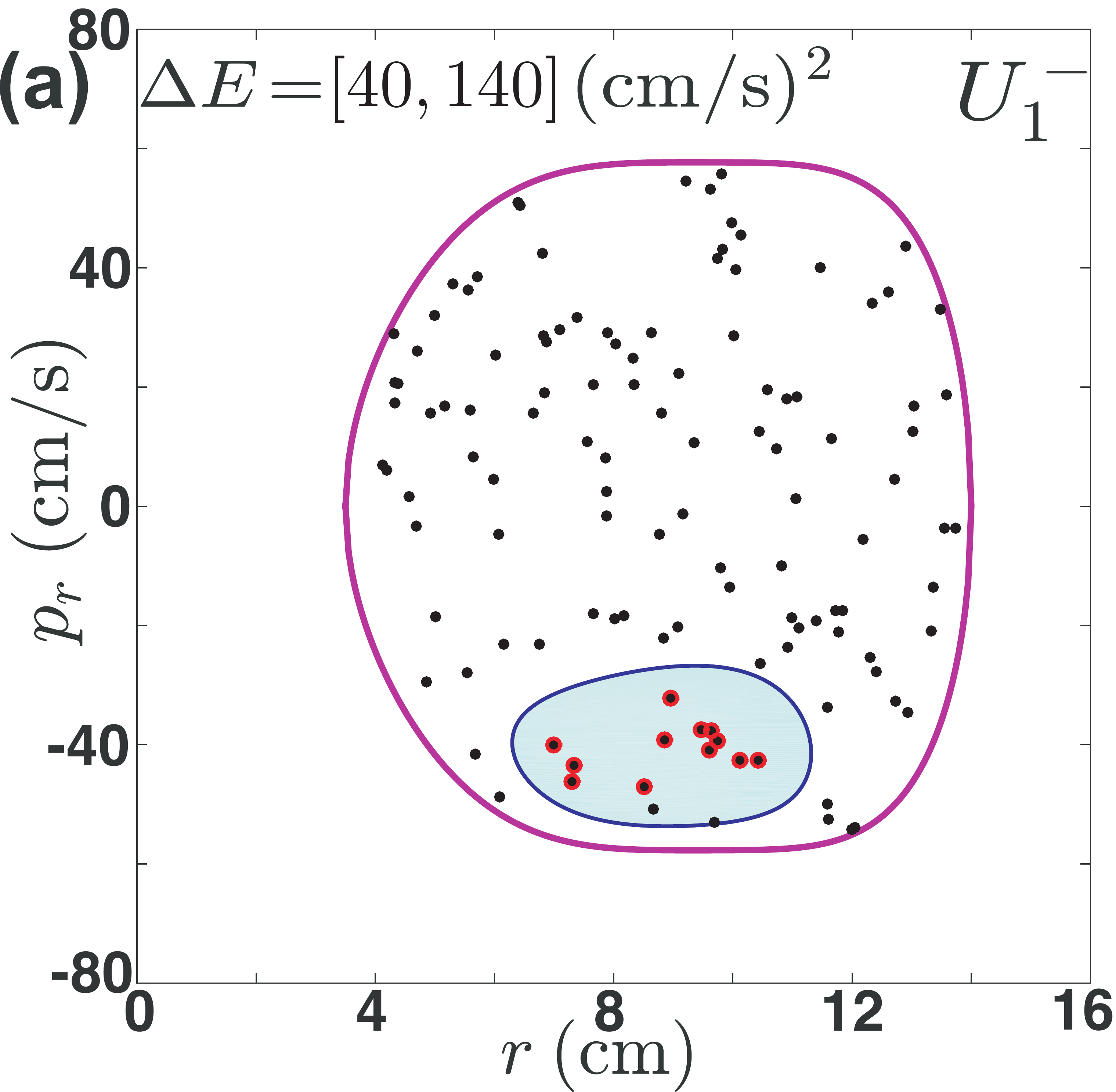}
\includegraphics[height=1.6in]
{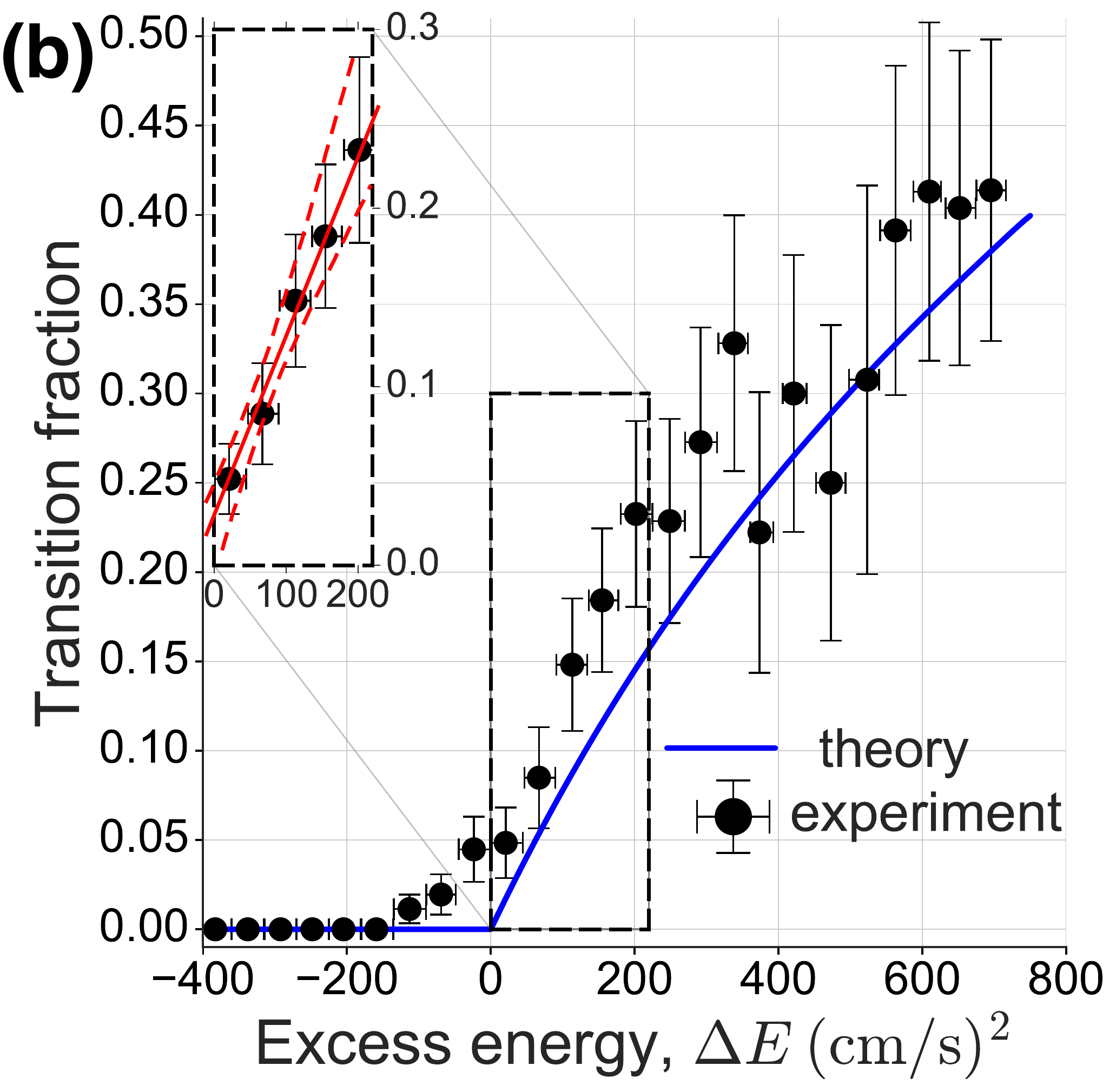}
\end{tabular}
\end{center}
\vspace{-5ex}
\caption{\footnotesize{\textbf{(a)} On the Poincar\'e section, $U_1^-$, we show 
a narrow range of energy 
($\Delta E \in (40,140)~{\rm (cm/s)}^2$) and label intersecting trajectories as 
no transition (black) and imminent transition (red) to quadrant 2, based on 
their measured behavior. 
The stable invariant manifold associated with the bottleneck periodic orbit 
at excess energy, $\Delta E = 140 \; {\rm (cm/s)^2}$, intersects the Poincar\'e 
section, $U_1^-$, along the blue curve. Its interior is shown in cyan and 
includes the experimental transition trajectories.
The outer closed curve (magenta) is the intersection of the boundary of the 
energy surface $\mathcal{M}(\Delta E)$ with $U_1^-$. 
\textbf{(b)} Transition fraction of trajectories as a function of excess 
energy above the saddle. The theoretical result is shown 
(blue curve) and experimental values are shown as filled circles (black) with 
error bars. For small excess energy above critical ($\Delta E = 0$), the 
transition fraction shows linear growth (see inset) with slope $1.0 \pm 0.23 
\times 10^{-3}$ (s/cm)$^2$ and shows  
agreement with the analytical result~\eqref{transit_frac}. A movie of 
increasing transition area on the Poincar\'e section, $U_1^-$, can be found 
\href{https://youtu.be/YZKYx0N9Zug}{here}.
}}
\label{fig:energy_poincare_section}
\vspace{-3ex}
\end{figure}
Grouping intersection points by energy, for example 
Fig.~\ref{fig:energy_poincare_section}(a), we get an experimental transition 
fraction, Fig.~\ref{fig:energy_poincare_section}(b), by dividing points which 
transitioned by the total in each energy range. Despite the experimental 
uncertainty from the image analysis, 
agreement between observed 
and predicted values is satisfactory. In fact, a linear fit of the experimental 
results for small excess energy gives a slope close to that predicted by 
\eqref{transit_frac} within the margin of error. Furthermore, the 
clustering of observed transitioning trajectories in each energy range, as in 
Fig.~\ref{fig:energy_poincare_section}(a), is consistent with the theory of 
tube dynamics. The predicted transition regions in each energy range account 
for more than 99\% of 
the observed transition trajectories.
\section{CONCLUSIONS}
We  considered a macroscopic 2 DOF experimental system showing transitions 
between potential wells and a dynamical systems theory of the 
conduits which mediate those 
transitions~\cite{KoLoMaRo2000,NaRo2017,ZhViRo2018}. The 
experimental validation presented here confirms the robustness of the conduits 
between multi-stable regions, even in the presence of non-conservative forces, 
providing a strong footing for predicting transitions in a wide range of 
physical systems. Given the fragility 
of other structures to dissipation (for example, KAM tori and periodic 
orbits), these phase space 
conduits of transition may be among the most robust features to be found in 
experiments of autonomous multi-degree of freedom systems. Furthermore, this
study lays the groundwork for experimental validation for $N = 3$
or more degrees of freedom system, such as ship dynamics~\cite{Mccue2005, *McCue2006}, 
buckling of beams \cite{ZhViRo2018} and geodesic lattice domes, hanging roller pins, isomerization and roaming 
reactions~\cite{Mauguiere2014,Bowman2017}.


~\\

\section*{Acknowledgments}
SDR and LNV thank the NSF for partially funding this work through grants 
1537349 
and 1537425.


\bibliography{ross_naik_refs,AIS}			

\bibliographystyle{aipnum4-1} 

\end{document}